# From Textual Experiments to Experimental Texts: Expressive Repetition in "Artificial Intelligence Literature"

Zhu Tianhua

Zhu Tianhua, Assistant Research Professor, Institute of Literature, SASS
E-mail: zhuth@sass.org.cn

**Abstract:** Since the birth of artificial intelligence 70 years ago, attempts at literary "creation" with computers are present in the course of technological development, creating what one might call "artificial intelligence literature" (AI literature). Evolving from "textual experiments" conducted by technologists to "experimental texts" that explore the possibilities of conceptions of literature, AI literature integrates primitive problems including machine thinking, text generation, and machine creativity, which exhibits the two-way interaction between social ideas and technology. In the early stage, the mutual support between technological path and artistic ideas turned out to be a failure, while AI-driven expressive repetitions are made probable in the contemporary technological context, paving the way for the transformation of AI literature from proof for technical possibilities to self-verification of literary value.

A. M. Turing's paper "Computing Machinery and Intelligence," originally published in 1950, marks the formal birth of AI technology. Over the past 70 years, the development of AI technology has led to continuous expansion of its application. Speaking of literature, AI reveals itself not only referentially in the narrative of science fiction (Wang 128), but also substantially in its intervention in texts. Starting from Christopher Strachey's "Love Letter" program in 1952, AI has involved itself in and penetrated into the generation of multiple genres of texts, notably poems and novels, to different degrees at various phases of its development. In recent years, AI system can even generate quantized recommendation of literary works, which has won attention in the field of literary criticism. In summary, the application of AI technology in literature has spawned "AI literature" in general.

Admittedly, "AI literature" is still only a tentative form in practice today. AI-generated texts differ greatly from human works, and the source of its value, the criteria for judgment, and even the object for such evaluation are also constantly changing. The ambiguity in value makes "AI literature" face many conceptual controversies, against which "AI literature" is unable to justify itself as "literature." In this regard, it is necessary to return to the specific context of both concept changes and technological development of AI, so as to provide a more reliable basis for theoretical inquiry to the characteristics of the practice and the logic behind the development of "AI literature."

I. **Practice Trajectory from "Textual Experiments" to "Experimental Texts"**

"AI literature" began with "textual experiments" conducted by technologists, with focus on computer's capability of handling words. In 1948, the world's first stored program electronic computer was built at the University of Manchester in the UK. Turing noticed the project instantly and became the deputy director of the computer laboratory at the University the following year, responsible for its software part. During the same while, Turing prepared and published his work on "Computing Machinery and Intelligence." In 1951, he invited Christopher Strachey to join his research team and appointed the latter to develop AI programs. Strachey completed the "Love Letter" program the following year, which is the earliest known text generation program (Link 55-56). However, Christopher Strachey's influence was bounded at campus. His pioneering works on AI have long been buried in the piles of old papers, until they were rediscovered by German artist David Link decades later. In comparison, another program called "Stochastic Texts," independently programmed by German engineer Theo Lutz in 1959, is a more well-known work of the same sort. The texts it generated were widely distributed, making it a recognized beginning for "digital poetry" (Funkhouser 37).

Both "Love Letter" and "Stochastic Texts" generate texts on the basis of randomness. The programs choose from a pre-defined list of rules and vocabulary to complete the "article" or "poem." Since it is a mere coincidence for computers to generate a fluent text, it is the artist's responsibility to choose from the results after dozens or even hundreds of runs, which has been a common practice at the early stage of "AI literature." In 1968, Institute of Contemporary Arts at London entitled an exhibition it held dedicated for application of computers in arts as "Cybernetic Serendipity," which was a very appropriate phrase to summarize the situation. Although the curators of "Cybernetic Serendipity" believed that the exhibition could show "artists' involvement with science, and the scientists' involvement with the arts" (Reichardt 5), and admittedly, a few painters and composers did participate, technologists still dominated the exhibition, as well as the overall situation of early "AI literature." Practitioners at the time were more indulged in exploring technical possibilities of generating and processing texts with computers, rather than the possible cultural implications or literary values of computer-generated texts. The "serendipitous" texts served only as evidences for technical viability in this case. A further comparison with the field of then-emerging computer graphics may help us to see how "indifferent" the pioneers of "AI literature" were to artistic values. A. Michael Noll, one of the pioneers who programmed computers to draw abstract curves, deliberately compared the outputs to formal characteristics of modernist paintings, and accordingly name his works as "computer art." Nonetheless, Lutz and his counterparts clearly expressed their exclusive interests in mathematics and technology, regarding computer-generated texts as the by-product of their textual experiments, showing hardly any intention to pursue "literary works."

Nevertheless, "Cybernetics Serendipity" marked an important step in transforming computer-generated texts to "literary works," especially "poems." Max Bense, German philosopher and one of the curators for the exhibition, presented systematically a "technicalized" framework of aesthetic judgments in his book *Introduction to Information Theory Aesthetics* in 1964. In a series of subsequent theoretical interpretations, he also put forward slogans such as "cybernetic and material poetry" and "random and topological poetry" ("Zur Lage" 166), calling on poets, artists and technicians to digitize the "perceptual structure" from the four aspects of "the semiotic, the metric (formal and structural organization), the statistical and the topological," and to establish a series of operable steps for writing poems ("Projekte" 11). From a contemporary point of view, "information theory aesthetics" is a doomed attempt (Nake 65). However, the goal-setting and the methodology for artistic creation Bense had advocated endowed significance to the use of computers in arts. His "information theory aesthetics" incented the attempts to implant technical experiment into artistic context, and "Cybernetics Serendipity" was an epitome.

The various cases shown in "Cybernetics Serendipity" provided vivid illustrations for Bense's artistic proposals, created broad impact, and changed the way people interpret "AI literature." For instance, some commentators noticed the fact that Lutz's chose from Franz Kafka's *Castle* to form the vocabulary in his "Stochastic Texts," and thus connected cybernetics with the idea of "social control" in the novel (d'Ambrosio 54). Some, represented by Bense himself, believed that the randomness of computers is the digital counterpart of "inspiration," which will endow computers with "autonomy." Bense appealed that practitioners should strengthen the symbolic linkages between technical matters that generate text and existing concept of cultural values. Under the framework of these linkages, the practitioners could confer meanings to the texts of "AI literature," instantiating what is called "manifesto-driven art" by Danto (33).

It is worth noting that applying technical matters to symbolize cultural concepts is consistent with the technical path of symbolistic AI (or Good Old-Fashioned AI, GOFAI). Guided by the philosophical view that "thoughts are symbolic" (Haugeland 93), the exploration of AI technology at that time was mainly centered on symbolic representation of rules and matters. As "AI literature" processes texts written in natural languages, a symbolic system computers and human beings share, it may considerably provide the most intuitive way of realizing AI. With the mediation of artistic ideas in "information theory aesthetics," the technological path of AI and "AI literature" could form a relation of mutual support and mutual proof, which propelled the establishment of its cultural connotation and artistic values.

However, two "AI winters" during the late 1970s and 1980s showed that the promises of GOFAI could not be achieved, the closed-loop of mutual support between technological path and artistic ideas broke up. On the other hand, the language model established on the basis of statistics and probability has made great progress since then, forming a competing substitute for rigid templates and rules in GOFAI. "AI literature" went through a paralleled period of silence and returned afresh. In 2018, when Microsoft Software Technology Center (Asia) (STCA) published the topical "collection of poems" titled *Sunlight Lost the Glass Window*, it became vain efforts for literary critics to crack down the image of the "girl poet" fabricated by technical staff and to draw awareness to the "programmatic operation" behind the text. In the past, programs meant rigid and mechanical combination of words. But now, with renewed technologies, "Xiaoice" can produce "fluent article" and poems with diverse contents based on user-uploaded pictures. Controversy arises on how to judge these "poems." Likewise, a text interactive game named "AI Dungeon" became a hit in 2019. It does not only reproduce the "hypertext" experiments of "electronic novels" in the 1990s with new technology, but also succeeds in getting rid of rigid templates and rules. The program can "write" varied types of "adventure stories" with prompts from the player in realtime interactions. These two cases show that the critical strategies for templates and rules that were once effective for symbolistic GOFAI are no longer applicable to the texts generated by more modern paradigms of AI technology. In other words, the new "AI literature" under new technological conditions requires a new way of judgment.

In interacting with the state-of-the-art "AI literature," people are likely to notice the tint of literary experiment to a certain extent – texts generated by "Xiaoice" and "AI Dungeon" urge people to reflect on the boundary of modern poetry or adventure novels, and even to reflect on the concept of

"literature." With appropriate interpretations, AI-generated texts can serve the conceptual renewal of "literature," which implies that "AI literature" has changed from textual experiment – that initially testifies technical possibility – to experimental texts that presents the prospect of technology testifying for literary possibility, after a short period of mutual support between technological path and artistic ideas. This historical change of course in practice corresponds to the deflection of theories that supported "AI literature," that is, the shift of focus among three primitive problems.

**II. Shift of Focus Among Three Primitive Problems of "AI Literature"**

The primitive problems of machine thinking, text generation, and machine creativity, have accompanied the exploration and development of "AI literature." They reflect the two-way interaction between social ideas and technological practice. (1) The problem of machine thinking has promoted textual experiment invention and formed the source of "AI literature." (2) Technology-led practices have imposed impact on the literary field, thus stimulating critics' concerns about text generation. (3) Finally, the problem of machine creativity as the culmination of two primitive problems, in which the insoluble dimension of values gradually becomes an unneglectable key issue in "AI literature" at the stage of experimental texts.

Turing responded explicitly to the question "Can machines think?" at the beginning of his "Computing Machinery and Intelligence." He stated that (1) "intelligence" lacks a clear definition, and (2) it can be replaced by "imitation game," namely "Turing Test," where the judge and the machine, isolated from each other, conduct their "conversation" through a teletypewriter. "Intelligence" is granted when the judge recognize the other party as a normal human being (Turing 433). In the test, the decision of whether the "other party" has "intelligence" is made not by the designer but the judge who encounters it, thereby shifting the focus of problem solving from the connotation of "machine thinking" to the condition where the judgements regarding "intelligence" are made. The "imitation game" requires a computer system that can automatically generate text, and "Love Letter" program proves its technical possibility. Strachey noticed that, even a relatively simple program is convincing enough, to make people believe that the machine is "thinking" (Strachey 26). Similar to Turing's view, he also believed that human "thinking," though much more complicated than what "Love Letter" program could do, would ultimately be imitated through sophisticated program design. As a result, the primitive problem that originated AI technology – machine thinking – formed a close relation with "AI literature" since its very beginning.

Turing was the first to transform the problem of machine thinking into an operable field for technology, but he is not the first to raise the problem. His attitude towards machine thinking has a direct historical origin. During his studies at Cambridge, Ludwig Wittgenstein avowed "Can machines think?" being a meaningless question in class (see TS309 78)[①], which was a considerate conclusion for his thinking about machine thinking since 1930. In addition, Wittgenstein's claim that "thought is the use of symbols" (see MS108 201) has gained attention beyond philosophical circles, as it contributed to conceptual preparation for symbolist AI. Herbert A. Simon's opinion was an example. It was precisely this technological proposition that echoed the symbolistic view of thinking so that GOFAI can get recognized before it produced any applicable effects. If we look back further, as early as the 17th century, Hobbes described national sovereignty as an "artificial soul" in the opening chapter of his masterpiece *Leviathan*; in the 18th century, the philosopher La Mettrie utilized his view of "man is a machine" to explain mental activities of human beings; let alone Hephaestus's handmaidens in Greek mythology. Although these prototypes of "thinking machines" existed only rhetorically, they provided conceptual preparations for the realization of "thinking machines" to various degrees.

Since Turing's time, the problem of machine thinking has been repeatedly rewritten in parallel to technological development. In Strachey's view, this problem can be rewritten as "Can people write programs for 'thinking';" the technical team at Microsoft's STCA aims at "emotional accompany" and "AI creation" to enrich "machine thinking" with capabilities other than knowledge and reasoning. But what has not fundamentally changed is that the texts generated by programs have always been regarded as manifestations of "machine thinking" or "intelligence:" "the ability to produce ideas or artefacts that are new, surprising, and valuable — is the acme of human intelligence, and necessary for human-level AGI (Artificial General Intelligence)." (AI: Its Nature 67) Having that said, what may contribute to "new, surprising, and valuable" textual works remained vague, and these preset goals alone could not provide sufficient clues for an operable path. "AI literature" may contribute to solving the problem of machine thinking, but it is insufficient to defend its literary values.

Compared with computer visual artists who have been fond of borrowing and appropriation of modernist painting, early "AI literature" lacks not only consciousness of literary history, but also literary theoretical ambition. Text generation programs only rely on intuitive grasp of "poetry" by technologists, and its resulting texts were too mediocre to be presented to the refined. However, this also helped "AI literature" maintain its primitivity. As a "phenomenon and process that has not been simplified and processed by scholars" (Zheng 82), its vigorous development has attracted attention from the literary world and given birth to new primitive problems.

In 1967, the Italian writer Italo Calvino delivered a lecture entitled "Cybernetics and Ghosts." He pointed out that the machine operation defamiliarized creative writing, and casted new doubts to reasons and motivations for generation of literary texts in the past. In Calvino's view, generating texts through programs was a "combinatorial game" and therefore belonged to the category of "poetry," and more importantly, such process was mathematically precise and logically rigorous. Calvino pointed out that such a tight causal relationship was exactly what had been lacking in all previous theories of literary creation – "By what route is the soul or history or society or the subconscious transformed into a series of black lines on a white page? Even the most outstanding theories of aesthetics were silent on this point." (267-268)

Concepts including soul, society, and the subconscious usually serve as explanatory elements for the source of literary meaning and values in aesthetics, but they are incapable of dealing with the situation after machines and programs started intervening in text generation process, for programs being essentially soulless, neither conscious nor subconscious. Programs are impossible to understand the general characteristics of the story, not to mention participating social activities, thus invalidating existent explanatory routes. Hence, the essence of Calvino's text generation lays on the following question: What gives, and under what circumstances would it give, meaning to the text? We may call it "text generation problem." And his answer to this problem is quite audacious, especially when noticing that himself being a writer: "And so the author vanishes – that spoiled child of ignorance – to give place to a more thoughtful person, a person who will know that the author is a machine, and will know how this machine works." (269) The individual "author's" identity or singularity has been sublated and reduced to be "the product and the instrument of the writing process" (268); in contrast, "a more thoughtful person," that is, a reader, should replace the position of author to provide the moment for literary values through reading. In other words, granting of meaning and values to texts got separated from text-generating, and the right of decision for literary values is ceded to the reader.

Calvino's experience as a writer and his almost radical criticism of the existing theories both imply that the problem of text generation is also primitive. Starting from the horizon the problem has opened, the operation of machines and programs and writing activities of human writers are equated under the metaphor of "literature machine," which indicates rich aesthetic potential and technological implication. Unfortunately, text generation has long been regarded as a process of pure mechanical execution in AI, restraining Calvino's influence on the practice of "AI literature." Meanwhile, as Calvino evaded the technological details of actual text-generating machines and believed that the mere theoretical possibility of "literature machine" is powerful enough to "give rise to a series of unusual conjectures," while the actual machine is unnecessary to build (266). In this way, the due consideration on real, computer-based text generation was cut off by purely conceptual speculation. However, in any case, the text generation problem still reflected the efforts to grasp new technology from the perspective of literature, mirroring how Bense tried to reinterpret culture from the perspective of technology. The two-way interpretations integrated technological and literary practice, constituting an epitome for the interaction between cultural ideas and technological development in "AI literature."

The problem of machine thinking has stimulated the technological development of "AI literature," and the technological development gave birth to the problem of text generation. In the development of "AI literature," these two iconic primitive problems have triggered a third primitive problem – problem of machine creativity. The problem itself has a long history, but it was not until text-generating programs that the problem became urgent to be answered.

In 1842, Ada Lovelace, mathematician and daughter of the British poet Lord Byron, pointed out in a famous commentary on the design sketches of "Analytical Machine" that, a machine for symbolic operations, no matter how sophisticated it is, "has no pretensions whatever to originate anything. It can do whatever we know how to order it to perform." (Lovelace 722) Turing tried to refute Lovelace's thesis in theory, and text-generating programs explicated the contradiction between the mathematical grasp of the "machine" and the performance given by the actual machine in practice: the instructions constituting "Love Letter" program are definite and detailed, while the program show unexpected behavior to people in actual operation (Strachey 28). This instinctively overturns Lovelace's judgement, and the problem of machine creativity – "Can a machine create?" – emerged out of it.

Strachey believed that computer-generated texts could provide "raw material for new ideas" for people to make use of (29). He and his following practitioners of "AI literature," at the stage of textual experiments, suspended value judgments and regarded machine creativity as a special case of machine thinking. But such suspension cannot last long. When we grant something as a literary "work," we are not making a factual judgment, but an evaluation. Accordingly, as the textual experiments have fully proved technical possibility of text generation through machines, the value dimension became an unavoidable focus. The Stuttgart Group led by Bense sought for cultural correspondence for technological elements in order to establish new criteria for valuation, through the mutual support between technical path and artistic ideas. However, since the break-up of the closed-loop, there has been a long-term absence of values in responses to the machine creativity

problem. Values have become the Achilles' Heel that hinders further development of "AI literature."

In recent years, some theorists have tried to imitate Turing by making use of the judgment of text characteristics to bypass the value dilemma. Selmer Bringsjord et al. proposed "Lovelace Test" in 2001, whose ultimate goal is to make AI "create" stories indistinguishable from human creations. Mark O. Riedl shared a similar vision and proposed a more sophisticated "Lovelace Test Version 2.0" in 2014, extending from the genre of stories to multiple forms of arts, including painting and poetry. These researchers believed that allowing machines to "create" texts similar to human works would help people understand human creative activities "scientifically," which was the core value of practical activities such as "AI literature" ("Computer Model" 23). However, such "value" is still insufficient to respond to the concerns regarding literary values.

When summarizing the discussion on the problem of machine creativity, Margaret A. Boden pointed out whether AI can "create" depended eventually on the critics' ethical positions (*The Creative* 166). Computer-generated texts per se are not sufficient to prove or disprove. At the same time, as experimental texts prosper, the concept of "literature" is explored actively in more contemporary cases of "AI literature." As a product of intersection of machine thinking and text generation, the problem of machine creativity has replaced the machine thinking and has become the key to "AI literature" at the experimental text stage; and its core lies in how to establish literary values. The changes in the primitive problems of "AI literature" suggest that to respond to this current key problem, it is necessary to look at the process of text generation and the machine's role in such process from the perspective of readers, viewers, and participants other than "intelligence" and "creative subject."

### III. Expressive Repetition of AI Technology Implementation

The critical point of the trajectory from a textual experiment that testifies the technical possibilities to an experimental text that self-testifies the literary values lies in the situation where people are willing to recognize connection between "AI literature" and "genuine" literature. So long as the problem of machine creativity is the junction of machine thinking (from a technological perspective) and text generation (from a literary perspective), the solution to the dilemma may also be achieved from two aspects: the technological particularity of AI, and the Calvionian "machine" in literature.

Based on the technological particularity, the technical restraint on AI is observed. From such observation, it is noteworthy that the situational requirements have been embedded in AI from the very beginning. The teletypewriter in Turing's design of "imitation game" was not only an interface between human and machine, but also represented an important situational setting. Considering that, in the era of Turing, only a few high-level research institutions had computers, and consequentially only technologists who participated in designing those machines could have access to them. In order to reduce preconceived influence of design perspective and force participants to pay attention only to the response made by the "other party," Turing designed a laboratory situation to judge "intelligence," in which the human-computer interface was an important part. However, since 1980s, multimedia and network communication technology has developed rapidly, rendering more humanized appearances for computer systems, and thus changing the situational presuppositions for judging "intelligence." As an increasing number of non-professionals have gained access to computers, the understanding that computers are the product of design no longer plays a major role in the situation of people's daily contact and interaction with them. The problem of machine thinking has retreated into background, and there is no necessity to dive into technical details and discuss from a designer's perspective . People regard almost unconsciously computers as intelligent agents, admitting that those machines have their own "logic" and "needs," and act "automatically" or even "autonomously" – that is, people divert from what Daniel Dennett called the "design stance" to the "intention stance" (87-90).

The boundary of "intelligence" in question changes with the situation where "intelligence" is perceived. People imagine "super intelligence," or face "animal intelligence," or interact with "machine intelligence;" in every situation, the scope of "intelligence" goes far beyond human intelligence as previous technologists pursued. In fact, everyone who know how a cat, a bird, or a sweeping robot acts can identify some certain extent of "intelligence" that those animals and machines possess, even they are unable to speak, compute, or create arts in the same way as humans do. The identification of "intelligence" is made through the perceptible presentations (actions, sounds, images, text, etc.) these agents present, and the connections between such presentations in different situations. In consequence, the situation in which intelligence is endowed goes beyond the laboratory, but the principle of "imitation game" – "judging from the behaviors in a given situation" has penetrated into every aspect of everyday life, and "intelligence" is recognized dynamically as a result of interaction.

Human intelligence is not the only form of intelligence, and creation is not the only embodiment of intelligence in art either. The text generation and the meaning endowment are not synchronized, and who "creates" is no longer a main factor that affects its meaning and values. For example, in performing arts, the dancer's behavior surely contains comprehensible contents. The association between the acting object and cultural values and meaning is not limited to the creating level. Strachey made use of this fact to say a program is to a computer what a typist is to a typewriter, and a pianist is to a piano (Strachey 25). Similarly, we could say that AI systems act as performers for specific literary activities, and they are to literature what a pianist is to music.

Strachey's analogy can serve as a reference for valuating technology from the perspective of art. However, the status of a pianist in music art is generally dependent upon the situation of concerts. Contrarily, the supposed "performer" of literary activities seemingly lacks a clear corresponding situation, especially in modern writer's literature. How can this reference point be effectively applied to literature? In this regard, we need to bear in mind Calvino's reminder that the performers of literary activities must be found in a more original form – "The narrator's voice in the daily tribal assemblies is not enough to relate the myth. One needs special times and places, exclusive meetings; the words alone are not enough, and we need a whole series of signs with many meanings, which is to say a rite." (271-272) In fact, anthropological, archaeological, and historical researches focusing on the formula of oral literature have systematically revealed that rituals were the situational requirements for early human 'text generation' to happen, and they were also important opportunities for texts to gain meanings. Similarly, the situational requirements of AI will provide the fundamental "times and places" for realization of artistic value of the "AI literature;" the repetitive operation of AI system is the basic way for it to gain meaning.

In the natural world and human social life, countless matters reflect repetition. Of course, the value of "AI literature" does not lie in the similar text forms brought about by the repetitive text generation process. On the contrary, the early textual experiments were bounded in "combinatorial games" and the generated text obviously embodied characteristics of mechanical repetition, making it unacceptable as literature. However, the latest developments in "AI literature" have changed such situation. For example, people can recognize distinctive features of modern poems from language style and chanting content of "Xiaoice." These "poems" are not stemmed from pre-defined rules and templates, and sometimes are unidentifiable for its direct origin in wording. In other words, although the repetitive text generation in new "AI literature" still relies on programs, it shows a very different appearance from mechanical repetition. To comprehend AI-generated "poems", one does not have to decide whether "Xiaoice" can or should be regarded as a qualified "imitator" of modern poets, or whether it has "subjectivity" or is engaged in "imitation." All of these doubts are irrelevant in terms of grasping the meaning of the text generation process, which declares clearly the characteristics and language style of modern poetry.

Texts in new type "AI literature" function in a way quite similar to to those in rituals. In rituals, the specific "texts" are only secondary and minor to the performance; they are not so much the individual "original" of the narrator but rather the props for conveying the cultural concepts as required by rituals. This is exactly the essence of what some scholars called "expressive repetition" (Zeng 90): Repetition in rituals is limited in time and space, namely it is situational; in addition, it maintains and conveys beliefs and values shared by a community, meaning that the repetition is collective. But just like a Shaman performing rituals for a clan is not necessarily a member of the clan, the performer of expressive repetition could be separated from the meaning subject implied in the expression. Furthermore, the overall consideration of the repetition process plays a vital role, much more than each specific repetitive behavioral representation, such as ritual behaviors or individual text narrated. Situationality, collectiveness, performer divisibility, and wholeness constitute four main characteristics of expressive repetition.

Examining the new type of "AI literature" practice represented by "Xiaoice" and comparing it with the mechanical repetition of old "AI literature," we can find that it also reflects four main characteristics mentioned above. First, as a "poem writing" system, "Xiaoice" achieves its design purpose as an "AI" system only when people recognize the text it generates as poems. The contextual requirements of AI technology are here to conform with the conventionality of literary situation. Given that "Xiaoice" is to bring the poetry first formulated a century ago came to the public under modern social circumstances, the old literary form not only obtains a dynamic presentation, but also extends the situation in which literary texts such as modern poems show up (people do not just read it in books or static web pages). This is where its situationality lies.

Second, "AI literature" at the experimental text stage is giving collective expression. "Xiaoice" covers common features of modern poems in expression form, and behind it is the modern poem text carefully selected for "training" by the designer. The selection of this corpus, together with the decision to make the results available to the public, reflects the designer's acceptance and recognition of modern poetry. Readers who are able to grasp the meaning of these lines of poem need to have considerable understanding of modern poetry. With respect to the inclination that technical implementations tend to generality, the intersection of readers (users) and designers in cultural concepts and values is more decisive for a system like "Xiaoice" to truly become "AI." Whether the operation of "Xiaoice" is appropriate does not depend on how "fluent" certain texts it generates are, but on whether it indeed expresses a set of literary concepts about "modern poems" shared by all these literary participants, including technologists building the system, readers who interact with the system with texts as interfaces, and even the numerous authors, readers, editors, and copyists of poetry anthologies in modern Chinese history of literature, etc. That is to say, every text generated by "Xiaoice" constitutes an example of "modern poems." In the continuous production and accumulation of these examples, the collectiveness of poets in the history of modern Chinese literature is presented, and people's identification with modern poetry is expressed.

Third, "Xiaoice" itself is only a specific computer system, and it may not have subjectivity. That is to say, the text generation in "AI literature" is divisible in implementation. Taking advantage of the text generated by "Xiaoice," people pay tribute to the sages of modern poetry; it does not matter whether "Xiaoice" or "self" has subjectivity or expresses something. Of course, although the performer of expressive repetition is different from the subject of expression, its content follows the collective literary concept-as a "narrator" or "executor," "Xiaoice" is to serve that literary concept for people to greet modern poetry. The reason for such phenomenon can only be found by returning to situationality, namely, in this combination of human and computer, in the repetitive text generation process, the literary contextual conventions play a role.

Finally, AI literature at the stage of experimental texts requires a holistic vision. Its wholeness is constituted of two aspects. On the one hand, this holistic vision is an overview of "repetitions" – all texts the system generates. Through the continuous "sampling" by "Xiaoice", people will discover the untouched expression of modern poetry and intuitively perceive its limitations. The text generation of AI literature presents literary concepts and value pedigree carried by "modern poems" in history, exploring its possible boundaries. In this sense, "AI literature" represented by "Xiaoice" constitutes literary experiments, and the computer-generated texts are experimental, hinting possible contribution of AI to literature. It urges people to rethink profoundly the concept of "modern poetry" and even "literature." When the AI system repeatedly generates literary styles, images, and content embodied in the "training corpus," it can be connected with a wider range of literary and cultural concepts, and the latter provides it with a basis for valuation. On the other hand, in a reader's view, there is nothing other than the text generation process performed by the machine. In the context of literature and arts, one interacts with computer systems more as a whole than as a mass of objects for technical analysis. If AI literature must achieve its own set-goal, the design stance must be put aside, to allow the viewers or participants to start from the most superficial overall appearance to conduct meaning processing and value judgment; that is also where aesthetics begins.

"AI literature" at the experimental text stage has embodied the four above-mentioned characteristics, implying that expressive repetition functions in contemporary AI literature through AI technology. The experimental texts are the direct result of expressive repetition, and this process itself makes "AI literature" one of the ways people express their shared literary ideas. While providing explanations for specific cases such as "Xiaoice," the expressive repetition performed by AI technology also appears as a way for "AI literature" to establish its own literary value.

**Conclusion**

In 1961, Shen Congwen (1902-1988), a well-known Chinese writer, predicted in his *Abstract Lyricism* that "the process of literature and arts creation has its generality, which can be controlled by powerful social forces, and can even be generated by electronic computers at another time." (533) Nicolas Schöffer, one of the first artists to use computer for artistic creation, worded his belief in a more eye-catching manner. He put that art with computers involved would surely replace or surpass human art and reduce "physical training and cultural accumulation" that were necessary for artistic training, and it almost negates human artistic practice (Кассйль 4). Such belief was obviously driven by the same logic as today's prevailing criticism that AI would put literature and arts into a crisis. However, the 70-year history of "AI literature" shows that such logic is not tenable. "AI" can only be "intelligent" when recognized by people in practice, within social and cultural conventions; as the performer, AI systems are uncapable of canceling creation, programming, and interpretations, needless to say "replacing" human beings.

Heidegger once said that in the face of "dangers" that "technology" brings to human beings, "art" may save them (954). It is true that at the moment when AI is applied practically, all kinds of controversies focusing on it are not uncommon. But a large part of it is connected with the dominant capital and power behind it, and it is a Luddistic mistake to take the negative externality for the inevitable result of technology. For instance, Bernard Stiegler believed that technological developments such as "big data" make people lose knowledge and commit themselves to "mindless" automatic machines, (117) while the root of such dystopian envision is the deprivation of people by capitalism so that people are no longer encouraged to pursue theoretical creation. The philosopher entrusted the important mission of saving the "noetic soul" to art. But the history of "AI literature" brings us to the opposite side that what art "saves" is not soul without technology; rather, it is the original visage of technology as the human essential forces that may "save" us: in the situations created in the name of arts and literature, the temporality, partiality, and contextual relevance that has been closely integrated with specific application of technologies get subverted, so that technology may at least temporarily breaks away from the chains of business-capital logic and becomes a way of expression by a multitude.

Literature is not enforced to face "AI" for pure technological reasons, and technology cannot prime new literary forms "automatically." Even if the clash of technology into literary creation seem to reach its climax, "AI literature" shows that the interlinkage of technology and art is rooted in the context of social culture all way through its development, paving a way cracking the antinomy of culture and technology. In short, AI as technology was born in a specific social and historical context. The progress of AI in generating texts has embodied social motivations with its own forms of representations. The transformation from "textual experiments" to "experimental texts" is achieved through expressive repetition, thus realizing the situational mutuality of literature conventions and "AI".

AI-related literary practice is in the ascendant. As a well-deserved pioneer among them, the primitive problems encountered in the "AI literature" development, the expressive repetition through technical means, and the focus shift from the creator "subject" to the viewers provide beneficial reference for appropriate understanding of the emerging literary forms.

## Notes

① When quoting Wittgenstein's *Nachlass*, this paper confirms to the numbering system formulated by G.H. von Wright, with manuscripts numbered with MS, the typescript by TS, and the subsequent page number is that of the facsimile.